\RequirePackage{fix-cm}
\documentclass[natbib,twocolumn,final]{svjour3}      % twocolumn
\smartqed  % flush right qed marks, e.g. at end of proof
\usepackage{graphicx}
\usepackage{gnuplot-lua-tikz}
\usepackage{multicol}
\usepackage{multirow}
\usepackage{hhline}
\usepackage{mathtools}
\renewcommand{\tabcolsep}{0.2 cm}
\begin{document}

\title{Relevance of analytical Buckley-Leverett solution for immiscible oil displacement by various gases}

\titlerunning{Relevancy of Buckley-Leverett solution for immiscible oil displacement by gases}        % if too long for running head

\author{Aleksandr~Zhuravljov \and Zakhar~Lanetc}

\authorrunning{A.~Zhuravljov \and Z.~Lanetc} % if too long for running head

\institute{A.~Zhuravljov
               \at Institute of Physics and Technology, Tyumen State University, Tyumen, 625003, Russia\\
               \email{a.s.zhuravljov@gmail.com}
               \and
               Z.~Lanetc
               \at School of Petroleum Engineering, The University of New South Wales, Sydney, Australia\\
               \email{lanetszb@gmail.com}
}


\maketitle

\begin{abstract}
In order to generate the valid numerical simulation model, the sufficient amount of gathered data from the oil field is required. However, it is not always possible to acquire such data at the initial stage of project development. \cite{buckley1942mechanism} developed the analytical solution allowing to easily assess the oil displacement efficiency. One of the main assumptions of this model is incompressibility of oil and injected fluid. For slightly compressible water and oil such assumption is rational. However, that is not always the case when the gas is injected. This research aims to identify the conditions at which the usage of the incompressible gas model is appropriate. Likewise, the cases when the model of compressible gas is required are also evaluated. To accomplish the goals of this research, the comparative analysis between the injection of compressible and incompressible gases was undertaken using the numerical solution of the correspondent reservoir engineering problems. The validation of the numerical model was {performed} showing that it matches the analytical Buckley-Leverett solution. The findings of this research indicate that the relative and absolute density change with the pressure of the injected gas has the profound impact on the convergence between two models under consideration. With the increase in the injection pressure, the discrepancy between the models of compressible and incompressible gas raises for all the considered injection fluids  ($CO_2$, $CH_4$ and $N_2$). Due to a steep slope of ’density-pressure’ curve for $CO_2$ at low initial reservoir pressure, the incompressible model cannot accurately predict the oil displacement efficiency by this gas at any reasonable injection pressure. All one-dimensional (1D) results are also representative for two-dimensional simulations (2D). However, the mismatch between two models increases considerably for 2D simulation scenarios. This study might be beneficial for those considering or researching the possibility of immiscible gas flooding by various gases at the particular oil field. Knowing some basic reservoir and technological parameters, the presented results might be used as simple screening criteria allowing to estimate the relevance of analytical Buckley-Leverett solution. 
\keywords{Reservoir simulation \and Two-phase flow \and Numerical methods \and Buckley-Leverett solution \and Immiscible gas injection}
\end{abstract}

\section{Introduction}
The extraction of hydrocarbons from oil and gas reservoirs is a complex process which depends on fluid and rock properties as well as on reservoir driving mechanisms. Primary oil recovery, in which natural reservoir energy is used to produce oil, is followed by the secondary stage where the reservoir pressure is typically maintained by water or gas flooding. Albeit, the injection of such fluids might commence from the beginning of the field life span due to technical and economic reasons \citep{ahmed2006reservoir}.

According to \cite{craig1971gas}, the gas injection projects can be traced to 1917, whereas \cite{Lake} reported the beginning of immiscible lean hydrocarbon gas flooding in the US from the 1930th. A number of gases are currently used for pressure maintenance. The most widespread among them are nitrogen ($N_2$), carbon dioxide ($CO_2$) and lean hydrocarbon gas (mainly $CH_4$) \citep{lake2014fundamentals}. The gas is typically injected into the overlying oil interval gas cap or into the oil column. In some cases $CO_2$ might be injected into deep saline aquifers in order to reduce greenhouse gas emissions \citep{kamali2017field, kimbrel2015experimental}. 

The fundamental manuscript ’Mechanism of  Fluid Displacement in Sands’ was presented by \cite{buckley1942mechanism}. This work represents the analytical solution of two-phase immiscible incompressible fluid transport through the porous medium allowing to estimate the dynamics of reservoir fluid displacement by the injectant. The proposed model was based on several assumptions. Among them are the incompressibility of all reservoir fluids, and the absence of capillary and compositional effects. The initial pressure and saturation in a reservoir have to be constant, and the injection flow rate has to be steady. 

The further development of Buckley-Leverett model was offered by \cite{welge1952simplified} who found the analytical solution for a saturation shock front. These studies assume that the oil displacement by injected gases can be estimated with sufficient accuracy, assuming gas incompressibility. This concept is also supported by several other research papers, such as \cite{kern1952displacement}, \cite{shreve1956gas}, \cite{pirson1977oil}, etc.
 
When the complexity of the oil reservoir does not correspond to the analytical Buckley-Leverett solution, it is more practical to use the numerical reservoir simulation techniques, allowing to implement complicated development conditions. However, in some cases express analyses are required, for instance, as an alternative for comparison with numerical simulation results. In the circumstances when the sufficient production data is not yet available, some methods allowing to estimate the oil displacement efficiency are vital. 

Importantly, there is an approach allowing to estimate the oil displacement by compressible gases analytically \citep{bedrikovetsky2013mathematical}. One chapter of the work conducted by Professor Bedrikovetsky scrutinises the effects of compressibility on two-phase displacement providing valuable findings.  However, the implemented analytical methods are significantly more sophisticated and, consequently, less appropriate for the first simple estimation of the oil recovery than the aforementioned Buckley-Leverett solution. Moreover, using such analytical models of the compressible gas injection, it is impossible to specify the classical boundary conditions, such as the constant injection rate and fixed initial reservoir pressure at the production well.

Thus, a comparative analysis of numerical compressible and incompressible gas injection models is essential. The aim of the paper is to evaluate at which circumstances the Buckley-Leverett solution, or incompressible gas numerical model generally, might be applicable for accurate estimation of the oil displacement efficiency. In order to scrutinize the research questions, the injection scenarios of most widely used gases will be simulated at different reservoir and injection pressures.  
 
Clearly, the numerical diffusion can to some extent affect the results of numerical calculations. If this is a case, it can be inappropriate to compare such numerical solutions with analytical models. Hence, the two numerical models accounting for gas compressibility and incompressibility are compared under the scope of this research, where the incompressible model represents the numerical analogue of the analytical Buckley-Leverett solution.

When the incompressible gas is considered, its density is held constant. If the average or maximum reservoir pressure is used for density estimation, the saturation profile depends on the distance between the production and injection wells. As this distance increments, the maximum and average reservoir pressures raise accordingly. Thus, as the first gas reached the bottom hole of the injection well, the density value for the incompressible case was fixed at the initial reservoir pressure. Such approach is also suitable as the reservoir pressure is constant around the production well in accordance with the Buckley-Leverett analytical model. 

The rest of the paper is structured as follows. Section \ref{sec:methods} describes the physical and mathematical representation of research objectives in the differential and numerical forms. Likewise, the geometrical characteristics of 1D and 2D models are also presented. Section \ref{sec:results} lists the results of the numerical analysis, while section \ref{sec:discussion} scrutinises and discusses the main findings of the research. Subsequently, the main conclusions are summarised in section \ref{sec:conclusion}. 

\section{Methods} \label{sec:methods}

A physical model represents immiscible transport of multiphase (oil and water) flow through porous media. The valuable assumptions include the absence of capillary and gravitational forces, incompressible oil, constant viscosity, as well as homogeneous and isotropic reservoir rock. The aforementioned phenomena can be described by the following system of equations~(\ref{eq:conserv_mass_1}~--~\ref{eq:boundary_cond}) \citep{muskat1937flow, barenblatt1960basic, dullien2012porous, bear2013dynamics}.
\begin{eqnarray}
\begin{gathered}
\label{eq:conserv_mass_1}
\int \limits_{V} \frac{\partial \phi \rho_{j} S_{j}}{\partial t} d V + \oint \limits_{\Omega} \phi  \rho_{j} S_{j} \vec{\upsilon}_{j} d\vec{\Omega} - \int \limits_{V} \tilde{q}_{j} d V = 0,
\end{gathered}
\end{eqnarray}
\vspace*{-5mm}
\begin{eqnarray}
\begin{gathered}
\label{eq:darcy}
\phi S_{j} \vec{\upsilon}_{j} = - \frac{k_{rj}\vec{\vec{k}}}{\mu_{j}} \vec{\nabla}P,
\end{gathered}
\end{eqnarray}
\vspace*{-4mm}
\begin{eqnarray}
\begin{gathered}
\label{eq:sum_sat_2ph}
S_{g}+S_{o} = 1,
\end{gathered}
\end{eqnarray}
\vspace*{-4mm}
\begin{eqnarray}
\begin{gathered}
\label{eq:initial_cond}
P\left(\vec{x}, 0\right) = P^{ini}, \; S\left(\vec{x}, 0\right) = S^{ini},
\end{gathered}
\end{eqnarray}
\vspace*{-6mm}
\begin{eqnarray}
\begin{gathered}
\label{eq:boundary_cond}
\phi S_{j} \vec{\upsilon}_{j} = 0 \; \text{on} \; \Gamma,
\end{gathered}
\end{eqnarray}
where $j=g\left(\text{gas}\right), o\left(\text{oil}\right)$, $\phi$~--~porosity, $\rho$~--~density, $S$~--~saturation, $\vec{\upsilon}$~--~velocity, $\tilde{q}$~--~density of source mass flow, $k_{r}$~--~relative permeability, $\vec{\vec{k}}$~--~absolute permeability, $\mu$~--~dynamic viscosity, $P$~--~reservoir pressure, $t$~--~time, $V$~--~volume, $\Omega$~--~surface. The formulae (\ref{eq:conserv_mass_1}) and (\ref{eq:darcy}) represent the law of conservation of mass and Darcy's equation, respectively. The algebraic expressions (\ref{eq:initial_cond}) and (\ref{eq:boundary_cond}) describe the applied initial and boundary conditions.

When the numerical solution is applied, the variable splitting ($P$ and $S_{g}$) is suitable and  it is convenient to convert the equations~(\ref{eq:conserv_mass_1}) and (\ref{eq:darcy}) into the following forms \citep{chen2006computational}:
\begin{eqnarray}
\begin{gathered}
\label{eq:conserv_P}
\frac{1}{\rho_{g}}\int \limits_{V} \phi \rho'_{g}  S_{g}  \frac{\partial P}{\partial t}d V  - \\
- \sum_{j}\frac{1}{\rho_{j}}\oint \limits_{\Omega} \vec{\vec{k}} \rho_{j}\frac{k_{rj}}{\mu_{j}} \vec{\nabla}P d\vec{\Omega}
- \sum_{j}\frac{1}{\rho_{j}}\int \limits_{V} \tilde{q} _{j}d V = 0,
\end{gathered}
\end{eqnarray}
\vspace*{-4mm}
\begin{eqnarray}
\begin{gathered}
\label{eq:conserv_S}
\int \limits_{V} \phi \frac{\partial \rho_{g} S_{g}}{\partial t} d V - \oint \limits_{\Omega} \vec{\vec{k}} \rho_{g} \frac{k_{rg}}{\mu_{g}} \vec{\nabla}P d\vec{\Omega} - \int \limits_{V} \tilde{q}_{g} d V = 0.
\end{gathered}
\end{eqnarray}

The model of a block-centred geometry is used for finite difference representation of the integral equations~(\ref{eq:conserv_P}) and (\ref{eq:conserv_S}) \citep{kazemi1976numerical, aziz1979petroleum, jamal2006petroleum, chen2006computational, smith1985numerical, ames2014numerical}:
\begin{eqnarray}
\begin{gathered}
\label{eq:conserv_P_numerical}
\alpha_{P} \Delta_{P}^{t}  + \sum_{j}\sum_{\Delta\Omega} \beta_{j}\nabla_{P} + \sum_{j}\frac{q _{j}}{\rho^{n}_{j}}= 0, 
\end{gathered}
\end{eqnarray}
\vspace*{-3mm}
\begin{eqnarray}
\begin{gathered}
\label{eq:conserv_S_numerical}
\alpha_{S} \Delta_{\rho S}^{t} + \sum_{\Delta\Omega} \gamma\overline{\Delta}_{S}^{t} + \sum_{\Delta\Omega} \delta + q_{g} = 0,
\end{gathered}
\end{eqnarray}
where $q$~-~mass flow rate. Operator~$\sum_{\Delta\Omega}$ denotes the summation over all surface elements $\Delta\Omega$ of a grid block. Other parameters which do not contain such operator describe the cell itself.  Boundary conditions~(\ref{eq:boundary_cond}) can be presented in the following form:
\begin{eqnarray}
\begin{gathered}
\label{eq:boundary_cond_numerical}
\nabla_{P} = 0 \; \text{on} \; \Gamma.
\end{gathered}
\end{eqnarray}

The applied numerical solution scheme refers to the Sequential solution method (SEQ-method) \citep{aziz1979petroleum}. The parameters containing unknown variables in the numerical model (\ref{eq:conserv_P_numerical}~--~\ref{eq:boundary_cond_numerical}) are given by (\ref{eq:operatorsP},~\ref{eq:operatorsS}):
\begin{eqnarray}
\begin{gathered}
\label{eq:operatorsP}
\Delta_{P}^{t} = P^{n+1}-P^{n}, \; \nabla_{P} = \frac{1}{L} \left( P_{+}^{n+1}-P_{-}^{n+1}\right),
\end{gathered}
\end{eqnarray}
\vspace*{-4mm}
\begin{eqnarray}
\begin{gathered}
\label{eq:operatorsS}
\Delta_{\rho S}^{t} = \rho^{n+1}_{g}S^{n+1}_{g}-\rho^{n}_{g}S^{n}_{g}, \; \Delta_{S}^{t} = S^{n+1}_{g}-S^{n}_{g},\\
\overline{\Delta}_{S}^{t}= \overline{S}^{n+1}_{g}-\overline{S}^{n}_{g},
\end{gathered}
\end{eqnarray}
where sign `$+$' or `$-$' as a subscript in the pressure symbol ($P_{\pm}$) means the orientation of the grid cell relative to the particular surface element $\Delta\Omega$. If the pressure subscript is `$+$', the direction of the particular grid cell is positive; otherwise, it becomes negative.

The averaging of any parameter, for instance $\overline{\rho}$, indicates arithmetic mean except the relative permeability $k_{r}$ for which the upstream weighting \citep{aziz1979petroleum} is used. Averaging is performed between two cells that share a common surface element $\Delta\Omega$.

The coefficients in the numerical model (\ref{eq:conserv_P_numerical}~--~\ref{eq:boundary_cond_numerical}) are given by equations (\ref{eq:parsP},~\ref{eq:parsS}):
\begin{eqnarray}
\begin{gathered}
\label{eq:parsP}
\alpha_{p} =  \frac{\rho'^{n}_{g}}{\rho^{n}_{g}} \frac{S^{n}_{g} \phi V }{\Delta t^{n+1}} , \; \beta_{j} = - \frac{k \overline{k}^{n}_{rj}}{\mu_{j}}  \Delta\Omega,
\end{gathered}
\end{eqnarray}
\vspace*{-4mm}
\begin{eqnarray}
\begin{gathered}
\label{eq:parsS}
\alpha_{S} = \frac{\phi V }{\Delta t^{n+1}} , \gamma = \overline{k'}^{n}_{rg}\varepsilon,  \;
\delta = \overline{k}^{n}_{rg}\varepsilon, \\
\varepsilon = - \overline{\rho}^{n+1}_{g} \frac{k}{\mu_{j}}\nabla_{P}\Delta\Omega
\end{gathered}
\end{eqnarray}

As the comparative analysis is undertaken between two numerical models (compressible and incompressible gases), it is critical to minimise the discrepancy between the analytical Buckley-Leverett model and its numerical analogous. As analytical Buckley-Leverett solution includes the conservation of initial reservoir pressure around the production well, the numerical model has to obey this condition (\ref{eq:P_cell_prod}).

Such condition might be obtained when analysing the equation derived by Buckley and Leverett in their model (Buckley and Leverett 1942). This equation does not contain reservoir pressure explicitly. However, using the Darcy and diffusivity equations, the appropriate condition for reservoir pressure might be obtained. Physically, the Buckley-Leverett model implies the immediate production of the fluid reaching the bottom hole of the production well. Clearly, this is only possible if the condition (15) is obeyed.

To the best of authors knowledge, the available and relevant reservoir engineering software does not allow to fix the reservoir pressure in the particular grid block. Therefore, all algorithms of numerical simulation were implemented using C++ programming language.
\begin{eqnarray}
\begin{gathered}
\label{eq:P_cell_prod}
P^{prod} = P^{ini}.
\end{gathered}
\end{eqnarray}

Production rates can be obtained by the following equations \citep{aziz1979petroleum} :
\begin{eqnarray}
\begin{gathered}
\label{eq:q_cell_prod}
q^{prod}_{total} =-\sum_{j}\sum_{\Delta\Omega} \overline{\rho}^{n}_{j}\beta_{j} \nabla_{P},\\
q^{prod}_{g} =\left(f'_{g}\Delta_{S}^{t}+f_{g}\right) q^{prod}_{total},
\end{gathered}
\end{eqnarray}
where $f_{g}=\frac{\frac{k_{rg}}{\mu_{g}}}{\frac{k_{rg}}{\mu_{g}}+\frac{k_{ro}}{\mu_{o}}}$~--~gas fractional flow.

In order to conduct the quantitative analysis, the defined 1D and 2D grid geometry and constant rock properties were selected and can be found in Table \ref{tab:geometry}.

\vspace*{-4mm}
\begin{table}[h]
\centering	
\footnotesize
\caption{Model characteristics}
\begin{center}
\label{tab:geometry}
\begin{tabular}{|c|c|}
\hline
\multicolumn{2}{|c|}{1D geometry}\\
\hline
Reservoir (X axis) & \textbf{$1400$} m\\
\hline
Cell (X axis) & \textbf{$0.7$} m\\
\hline
Number of cells & \textbf{$2000$} \\
\hline
\multicolumn{2}{|c|}{2D geometry}\\
\hline
Reservoir (X $\times$ Y axes) & \textbf{$1000 \times 1000$} m\\
\hline
Cell (X $\times$ Y axes) & \textbf{$20\times 20$} m\\
\hline
Number of cells & \textbf{$2500$}\\
\hline
\multicolumn{2}{|c|}{Rock properties}\\
\hline
Porosity & \textbf{$0.2$} \\
\hline
Permeability & \textbf{$1$} Darcy\\
\hline
\end{tabular}
\end{center}
\end{table}

\section{Results} \label{sec:results}
This section reveals the results obtained by using the numerical simulation model discussed above. Two sets of identical calculations were conducted: 1D and 2D immiscible gas displacement of oil. Subsequently, numerical models of compressible and incompressible gas injection scenarios were simulated. For both 1D and 2D cases, varying parameters included injected gas ($CO_2$, $CH_4$ and $N_2$), as well as initial and maximal reservoir pressures (see Table~\ref{tab:calc_scenarios}).

\begin{figure*}[ht!]
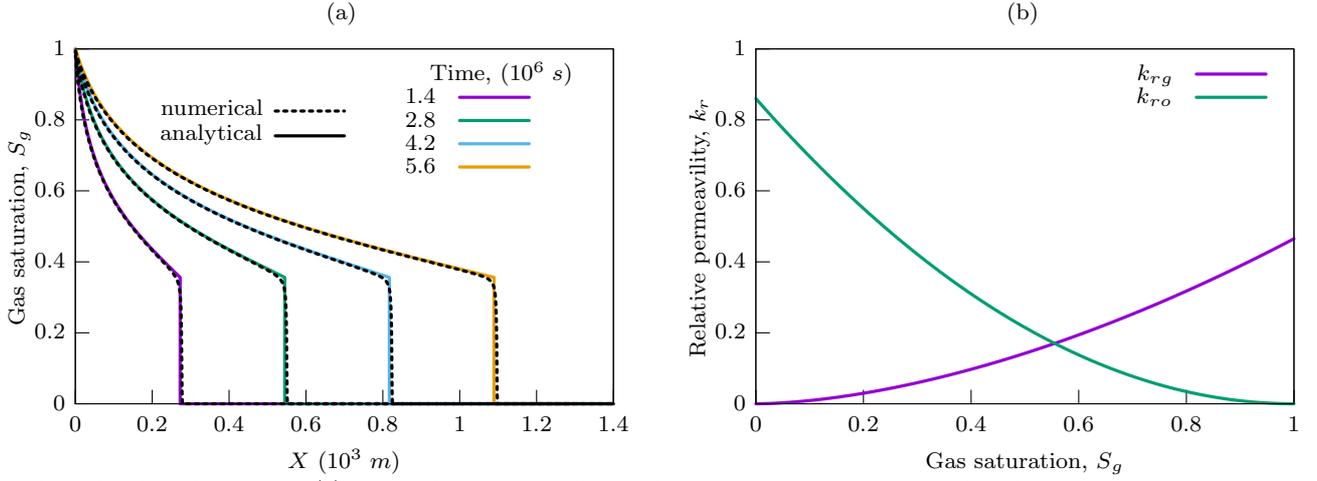

\centering
\footnotesize{\input 1_validation1D_CO_2_3.txt }
\vspace*{-6mm}\caption{Validation. 1D model. (a)~Saturation profiles. Comparison of numerical solution with Buckley-Leverett analytical model. (b)~Relative permeability curves used.}
\label{pic:1}
\end{figure*}

The validation of the applied numerical model was generated to ensure the convergence between the analytical Buckley-Leverett solution~\citep{buckley1942mechanism} and its analogous incompressible numerical model (see Fig.~\ref{pic:1}). The irreducible oil and gas saturations were absent for the validation scenario to exclude the numerical diffusion around the injection well. Such approach is correct as the validation case aims to prove the principal appropriateness of the numerical solution.

\begin{figure*}[ht!]
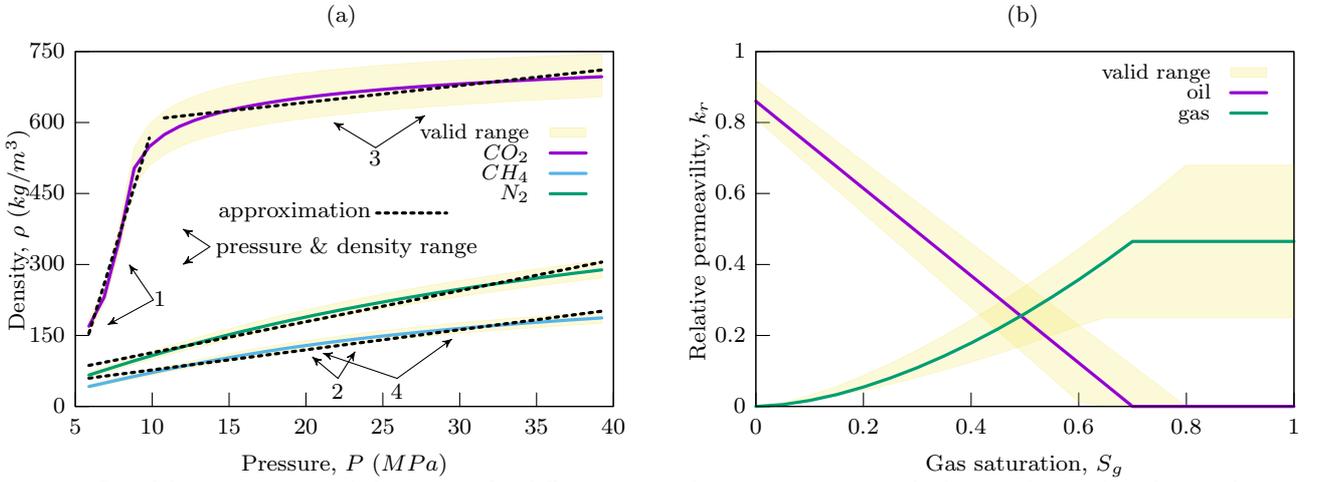

\centering
\footnotesize{\input 2_dens_rel_perm.txt }
\vspace*{-6mm}\caption{(a) Set of density curves and their range for different injected gases. 1, 2, 3, 4 -- calculations depicted in the article (see Table \ref{tab:calc_scenarios}). (b) Set of relative permeability curves and their range for oil-gas system.}
\label{pic:2}
\end{figure*}

Using Peng-Robinson equation of state \citep{peng1976new}, 'density vs. pressure' curves were built for all gases. This ensures that gas compressibility, and thus substantial density variations with pressure, are incorporated into the model (see Fig.~\ref{pic:2}a). Using equations of the straight line, 'density vs. pressure' relationships were further approximated for all the injected fluids. The linear approximation of density change with pressure was chosen as all the trend lines appropriately matched with approximated curves with insignificant errors. Likewise, such approach allowed to determine and clearly visualise two distinctive zones at the $CO_2$ 'density vs. pressure' curve.

The approximations are depicted in the form of dashed lines in Fig.~\ref{pic:2}a and listed in Table \ref{tab:calc_scenarios}. Knowing that the compressibility is a function of density, one can clearly see (Fig.~\ref{pic:2}a) how density of the particular gas changes with pressure when the compressible model is considered. Additionally, Corey correlation exponents \citep{corey1954interrelation} were taken from the research conducted by \cite{parvazdavani2013gas}, generalised and used as an input for numerical flow simulator for all gases under consideration (see Fig.~\ref{pic:2}b). 

\begin{table*}[ht!]
\centering	
\footnotesize
\caption{Calculation scenarios}
\begin{center}
\label{tab:calc_scenarios}
\begin{tabular}{|c|c|c|c|c|c|c|c|c|}
\hline
\multicolumn{2}{|c|}{Calculations depicted}& $1$ & $nd^*$ & $nd^*$ &	$2$ &	$3$ & $nd^*$	& $4$ \\
\hline
\multicolumn{2}{|c|}{Injectant gas} & $CO_{2}$ &	$CO_{2}$	 & $N_{2}$ &	$CH_{4}$ &	$CO_{2}$ &	$N_{2}$	& $CH_{4}$ \\
\hline
$\rho_g$ ($kg/m^3$) =  &$a$& $105.22$ &	$3.53$	 & $6.55$ &	$4.24$ &	$3.53$ &	$6.55$	& $4.24$ \\
\hhline{~--------}
= $a \cdot P \left( MPa \right) + b$ & $b$& $465.04$ &	$571.45$	 & $48.11$ &	$34.73$ &	$571.45$ &	$48.11$	& $34.73$ \\
\hline
\multicolumn{2}{|c|}{$P_{ini}$ ($MPa$)} & $6$ & \multicolumn{3}{|c|}{20} & \multicolumn{3}{|c|}{20}	 \\
\hline
\multicolumn{2}{|c|}{$P_{max}$ ($MPa$)} & $9$ & \multicolumn{3}{|c|}{23} & \multicolumn{3}{|c|}{35}		 \\
\hline
\multicolumn{2}{|c|}{$\mu_g$ ($mPa \cdot  s$)}& $0.02$ & $0.05$ & $0.03$ & $0.02$ & $0.05$ & $0.03$ & $0.03$ \\
\hline
\multicolumn{2}{|c|}{$\mu_o$ ($mPa \cdot  s$)} &$0.62$ & \multicolumn{3}{|c|}{$0.47$}	& \multicolumn{3}{|c|}{$	0.43$}		\\
\hline
\multirow{2}{*}{$q_g$ ($10^{3} kg \cdot s	$)} &1D& $0.50$ & $2.50$ & $0.70$ & $0.46$ & $13.80$ & $3.80$ & $2.60$ \\
\hhline{~--------}
&2D& $237$ & $920$ & $260$ & $165$ & $4930$ & $1390$ & $930$ \\
\hline
\multicolumn{9} {c} {$nd^*$ - not graphically depicted calculations, but their results are discussed.}  \\
\end{tabular} 
\end{center}
\end{table*}

\subsection*{1D calculations}
Figures \ref{pic:3}~--~\ref{pic:6} highlight several most demonstrative 1D results. The calculation scenarios were divided into several categories. 

Firstly, the $CO_2$ injection case was simulated at a low initial reservoir pressure and small injection pressure, due to a steep slope of $CO_2$ `density vs. pressure' curve in this range. For more details on pressure interval for this scenario see label `1' in Fig.~\ref{pic:2}a and in Table \ref{tab:calc_scenarios}. According to Fig.~\ref{pic:3}b, the significant gap between compressible and incompressible saturation fronts can be seen for such $CO_2$ injection case. Such difference between saturation fronts also affects the well flow rates, leading to the increase in accumulated production of oil, which is considerably higher for incompressible model (see Fig.~\ref{pic:3}a).

\begin{figure*}	[ht]
\centering	
\footnotesize{\input 3_compare1D_CO_2_1.txt }
\vspace*{-3mm}\caption{Injectant gas $CO_2$. 1D model. Scenario $1$. Comparison of compressible and incompressible cases. (a)~Production profiles. (b)~Saturation profiles. }
\label{pic:3}
\end{figure*}
\begin{figure*}[ht]
\centering	
\footnotesize{\input 4_compare1D_C_H_4_1.txt }
\vspace*{-3mm}\caption{Injectant gas $CH_4$. 1D model. Scenario $2$. Comparison of compressible and incompressible cases. (a)~Production profiles. (b)~Saturation profiles. }
\label{pic:4}
\end{figure*}

Secondly, the injection of three different gases ($CO_2$, $CH_4$ and $N_2$) was simulated at large initial reservoir pressure with small injection pressure (e.g. for $CH_4$ case see Fig.~\ref{pic:4}). For more details on pressure range for this calculation, see number `2' in Fig.~\ref{pic:2}a and in Table \ref{tab:calc_scenarios}. The obtained results provide the minor, although visible, gap between the compressible and incompressible saturation fronts for all of the injected gases. Subsequently, both well oil rates and accumulated production values indicated close compliance between compressible and incompressible models of all the injected gases. 

Finally, large injection pressure scenarios were simulated for all the gases at high initial reservoir pressure. For more details on pressure diapason for this scenario, see number `3' for $CO_2$ and `4' for $CH_4$ in Fig.~\ref{pic:2}a and in Table \ref{tab:calc_scenarios}. 
Carbon dioxide injection scenario even at large injection pressure showed proximity between the progress of incompressible and compressible immiscible fronts towards the production well. The same trend can be observed by looking at closely matching values of the well oil rate and its accumulated production (see Fig.~\ref{pic:5}a).

As can be seen from Fig.~\ref{pic:6}, the incompressible saturation front significantly outstrips the compressible saturation profile for the $CH_4$ case. Likewise, a substantial difference of well production rate and accumulated production profiles can also be observed, especially before the gas break through. Similar results were obtained for the not depicted $N_2$ immiscible injection case. 

\begin{figure*}[ht]
\centering
\footnotesize{\input 5_compare1D_CO_2_3.txt }
\vspace*{-3mm}\caption{Injectant gas $CO_2$. 1D model. Scenario $3$. Comparison of compressible and incompressible cases. (a)~Production profiles. (b)~Saturation profiles. }
\label{pic:5}
\end{figure*}

\begin{figure*}[ht]
\centering
\footnotesize{\input 6_compare1D_C_H_4_2.txt }
\vspace*{-3mm}\caption{Injectant gas $CH_4$. 1D model. Scenario $4$. Comparison of compressible and incompressible cases. (a)~Production profiles. (b)~Saturation profiles. }
\label{pic:6}
\end{figure*}

\subsection*{2D calculations}

As mentioned above, calculation scenarios for 2D geometry repeated 1D scenarios. For the sake of visibility, two out of seven 2D results are presented in this paper.

Figure~\ref{pic:7} represents the similar calculation scenario as was undertaken for 1D $CO_2$ (see Fig.~\ref{pic:3}) case at the low initial reservoir and injection pressures, while Fig.~\ref{pic:8} repeats the 1D $CH_4$ injection at the high initial reservoir and injection pressures. It is clear that substantial difference between the progress of compressible and incompressible gas fronts towards the production well was also preserved for 2D cases. Similarly, well oil rates and its accumulated productions were also considerably low for compressible cases in comparison with incompressible scenarios. Considering the 1D and 2D cases, one can observe the higher divergence between compressible and incompressible cases for 2D reservoir geometry.

\section{Discussion} \label{sec:discussion}
The two numerical models of immiscible oil displacement by gas injection have been compared. The first of them considered injected gas as the incompressible fluid mimicking the Buckley-Leverett model, while the second accounts for gas compressibility. Any outstanding characteristics identified during the numerical simulation process are discussed below.

\subsection*{Gas `density - pressure' relationship}
To consider gas compressibility, Peng-Robinson equation of state was utilised allowing the construction of `density vs. pressure' relationship for three most widely used for immiscible oil displacement gases, namely $CO_2$, $N_2$ and lean hydrocarbon gas ($CH_4$ is assumed). The range of considered reservoir pressures varied from 5 to 40 MPa which corresponds to current field practice in the oil and gas industry \citep{taber1997eor}. Examining the obtained `density vs. pressure' plot, several conclusions were made allowing to identify calculation scenarios.  

According to Fig.~\ref{pic:2}a, $CO_2$ has a steep density increase at the pressure range between 5 and 10 MPa. However, as the pressure approaches 10 MPa , $CO_2$ `density vs. pressure' curve flattens significantly, until the end of the analysed pressure range. The trend is different for $CH_4$ and $N_2$ cases, where the slope of `density vs. pressure' curves is fairly constant along the whole pressure interval. 

Thus, the assumption was made that both $CH_4$ and $N_2$ `density-pressure' relationships might be approximated by the equation of a straight line. In turn, two such equations are required for the $CO_2$ injection case. For more details see Table \ref{tab:calc_scenarios} and Fig.~\ref{pic:2}a. To construct `density vs. pressure' curves, geothermal gradient \citep{Lake} and hydrostatic water head data was used to calculate temperature and pressure values for different elevation depths. The results were utilised to obtain gas compressibility factor allowing to calculate gas density using an equation of state.

However, in the circumstances when primary recovery precedes the immiscible gas injection, reservoir pressure can be significantly reduced. In such case, reservoir temperature is still consistent with geothermal gradient. Thus, for the particular reservoir pressure, temperature values can vary, affecting gas density.  The possible density variation at particular pressure was also taken into consideration and does not affect the main qualitative results. Such variation range is depicted as yellow blurred diapason at Fig.~\ref{pic:2}a.

\subsection*{Relative permeability curves}
Relative permeability data was constructed using Corey correlation exponents taken from the research conducted by \cite{parvazdavani2013gas}. The decision to use this particular data was made on the basis of experimental focus of the paper, as well as the broad range of Corey exponents for reservoirs with different fluid and petrophysical characteristics. As indicated by \cite{ghoodjani2011experimental} and \cite{al2015capillarity}, the shape of relative permeability curves does not considerably change for various gases in a particular reservoir, instead the irreducible oil saturation values differ frequently. Such peculiarity allowed to construct generalised gas/oil relative permeability curves which are valid in the range depicted as yellow blurred area at Fig.~\ref{pic:2}a.

\begin{figure*}[ht]
\centering	
\footnotesize{\input 7_compare2D_CO_2_1_2D.txt }
\vspace*{-3mm}\caption{Injectant gas $CO_2$. 2D model. Scenario $1$. Comparison of compressible and incompressible cases. (a)~Production profile. (b)~Saturation profile for $4.5 \cdot 10^{7} s$ }
\label{pic:7}
\end{figure*} 

\begin{figure*}[ht]
\centering	
\footnotesize{\input 8_compare2D_C_H_4_2_2D.txt }
\vspace*{-3mm}\caption{Injectant gas $CH_4$. 2D model. Scenario $4$. Comparison of compressible and incompressible cases. (a)~Production profiles. (b)~Saturation profile for $5 \cdot 10^{6} s$ }
\label{pic:8}
\end{figure*} 

\subsection*{Injection scenario 1}
As can be seen from scenario 1 (see Table \ref{tab:calc_scenarios} and Fig.~\ref{pic:3}), $CO_2$ injection at a low initial reservoir pressure leads to a significant discrepancy between compressible and incompressible numerical models even at a low injection pressure. Such phenomenon is characterised by substantial absolute and relative change in $CO_2$ density at pressure range between 5 and 10 MPa (see `1' at Fig.~\ref{pic:2}a). 

Significant lag in the saturation progress of compressible gas front affects the well oil production profile. As  is evident from Fig.~3a, the well oil flow rate in the compressible case is lower, comparing to the incompressible numerical solution. We can state that flow rate is directly proportional to the velocity of the saturation front. The well oil flow rate is constant for incompressible case before the gas breaks through. Subsequently, the well oil flow rate drops after breakthrough, as the well produces both oil and gas.

The situation is different for compressible gas where immiscible gas front accelerates during the progress towards the production well. It should also be noted that flow rates almost become equal at the later stage, as several pore volumes were injected into the oil reservoir.

\subsection*{Injection scenario 2}
As was described in the results section, the second scenario implied the pressure maintenance by all three gases at the high initial reservoir and low injection pressures. Importantly, the $N_2$ and $CO_2$ injection scenarios were not depicted in this article. Former due to its close similarity to $CH_4$ case, and latter because of its perfect match between compressible and incompressible solutions. Thus, all the qualitative results obtained for $CH_4$ are also applicable to $N_2$. As can be seen from Fig.~\ref{pic:2}a, both absolute and relative changes in density are small for all the injected gases at the range between 20 and 23 MPa.  As a result, the compressible and incompressible solutions for such scenarios differ insignificantly, though some variations can be observed, for instance, in the case of  $CH_4$ injection (see Fig.~\ref{pic:4}). 

\subsection*{Injection scenarios 3 and 4}
Scenarios 3 and 4 (see Table \ref{tab:calc_scenarios}) simulate the injection of $CH_4$ and $CO_2$ at the high initial reservoir (20 MPa) and large differential (15 MPa) pressures (see Fig.~\ref{pic:6} and \ref{pic:4}, respectively). While $CO_2$ injection scenario still provides close matching between incompressible and compressible numerical models, this is not the case for $CH_4$ immiscible flooding. Such discrepancy occurred due to the difference in the slope for `density vs. pressure' curve for these gases at the considered pressure range. The relative and absolute changes in the density of $CO_2$ are still not sufficient to cause any significant difference between compressible and incompressible saturation fronts. For displacement of oil by $CH_4$ injection, saturation front for incompressible gas overrides the compressible model for more than 200 meters before the break through. The results of $N_2$ injection are highly consistent with the $CH_4$ case and are not presented graphically in this research.

\subsection*{Injection scenarios 1 and 4 (2D)}
Scenarios 1 and 4 were chosen for 2D simulation as the ones showing the highest discrepancy between the compressible and incompressible cases. Importantly,  the 2D calculations are qualitatively consistent with their 1D analogues. However, some refinements should be noted. There is a much larger mismatch between the compressible and incompressible simulations in terms of oil flow rates in 2D models in comparison with their 1D analogues. 

That clearly leads to the even higher gap between the compressible and incompressible saturation fronts. Thus, there is even larger difference between the accumulated productions of the two models. We can also conclude that mismatch between the compressible and incompressible models will increase even more significantly for 3D simulations.

\section{Conclusion} \label{sec:conclusion}
The concluding section revisits the research questions raised in the introductory part, summarizes the main results and findings of this paper, and provides implications and conclusion based on these findings. 

The main aim of this study was to explore the possible application of Buckley-Leverett model for the oil displacement by immiscible gases. The broader focus of this paper was on the comparison between the compressible and incompressible gas injection models in the oil fields.

In order to scrutinise the research questions, the numerical model was created allowing to repeat the analytical Buckley-Leverett solution, as well as to simulate the injection of compressible fluids. The `density vs. pressure' curves were built for the most widely used injection gases ($CO_2$, $CH_4$, $N_2$). Following the qualitative examination, these curves were used as an input parameter to the reservoir flow simulator. 

Conducted flow simulations have allowed to summarise the main research findings as follows. 

\begin{enumerate} 
\item The `density vs. pressure’ relationships for $CH_4$ and $N_2$ might be approximated using the equation of a straight line. However, two such equations are required to approximate $CO_2$ density due to its non-linear shape. 

\item The assumption that at high reservoir pressures the results from compressible and incompressible gas behaviour are similar, while there is a significant difference for lower reservoir pressures is frequently made looking expectable and trivial. However, as was revealed in the research, this may be or may not be correct depending on the injected gas or the reservoir pressure variation during the gas flooding process.

\item The injection of $CO_2$ at a low initial reservoir pressure cannot be correctly predicted by assuming the gas as incompressible fluid, even at low injection pressure. The discrepancy between compressible and incompressible cases increases significantly as injection pressure increments.

A similar trend can be observed when oil is displaced by $CH_4$ and $N_2$ at a high injection pressure. The initial reservoir pressure does not affect the results in this case. 

Hence, the reservoir performance cannot be accurately predicted using Buckley-Leverett solution, for the cases mentioned above.

\item The injection of $N_2$ and $CH_4$ at a low injection pressure and either high or low initial reservoir pressure leads to sufficient match between compressible and incompressible injection scenarios. Considering high initial reservoir pressure and low $CO_2$ injection pressure, perfect compliance between the two numerical solutions is observed. 

Additionally, pressure maintenance using $CO_2$ at high injection pressure does not affect significantly the convergence between compressible and incompressible models at high initial reservoir pressure. Albeit, the mismatch between the two models is higher in comparison with oil displacement by $CO_2$ at a low injection pressure. 

\item Both absolute and relative density change with pressure affects the results of the simulation. As these values increment, the difference between compressible and incompressible models raises respectively. Thus, these attributes should be taken into consideration when the immiscible gas injection is simulated using Buckley-Leverett model. 

\item The velocity of compressible gas saturation front increases in the reservoir as it progresses towards the production well. Such phenomenon is uncommon for the Buckley-Leverett solution where saturation front moves with constant velocity.

\item Two-dimensional simulation allows to qualitatively prove 1D results. However, the difference between compressible and incompressible solutions increases in the 2D case in comparison with 1D calculations.
\end{enumerate}

\section*{Acknowledgement} 
Authors would like to express the deepest appreciation to Dr.~Furqan  Hussain,
not to mention his valuable role in the idea of this research and unsurpassed knowledge. Without the support and
valuable advice of our friends, Denis Buxmann, Sergei Iskakov, and Andrey Kutuzov, it would not have been possible to complete this work. It should also be noted that  the C++ template library for linear algebra `Eigen' \citep{eigenweb} was very helpful and convenient for the solution of linear algebraic equations.

This is a pre-print of an article published in Journal of Petroleum Exploration and Production Technology. The final authenticated version is available online at: https://doi.org/10.1007/s13202-018-0516-6 .

\bibliographystyle{spbasic}          % basic style, author-year citations
\bibliography{mybibfile}

\end{document}